\documentclass[12pt]{article}
\usepackage{hyperref}
\usepackage[english]{babel}
\usepackage{amsmath}
\usepackage{latexsym}
\usepackage{amssymb}
\usepackage[all]{xy}
\usepackage{graphicx}
\usepackage{epsfig}
\usepackage{psfrag}

\usepackage{color,eucal}

\numberwithin{equation}{section} \numberwithin{table}{section}
\numberwithin{figure}{section}

\begin{document}

\begin{titlepage}
   \begin{flushright}
{\small}
  \end{flushright}

   \begin{center}

     \vspace{20mm}

     {\LARGE \bf Universal corner contributions \\
     \vspace{3mm}

     ~~~to entanglement negativity
     \vspace{3mm}}

     \vspace{10mm}
 Keun-Young Kim${}^{a}$, Chao Niu${}^{a}$ and Da-Wei Pang${}^{b}$\\
     \vspace{5mm}
      {\small \sl ${}^{a}$School of Physics and Chemistry, Gwangju Institute of Science and Technology, \\Gwangju 500-712,
Korea\\
      \sl ${}^{b}$Mathematical Sciences and STAG Research Centre, University of Southampton\\
      Southampton SO17 1BJ, UK}\\

     {\small \tt fortoe\underline{~}at\underline{~}gist.ac.kr,}
     {\small \tt chaoniu09\underline{~}at\underline{~}gmail.com,}
     {\small \tt d.pang\underline{~}at\underline{~}soton.ac.uk}

     \vspace{10mm}

   \end{center}

\begin{abstract}
\baselineskip=18pt
It has been realised that corners in entangling surfaces can induce new universal contributions to the entanglement entropy and R\'{e}nyi entropy. In this paper we study universal corner contributions to entanglement negativity in three- and four-dimensional CFTs using both field theory and holographic techniques. We focus on the quantity $\chi$ defined by the ratio of the universal part of the entanglement negativity over that of the entanglement entropy, which may characterise the amount of distillable entanglement. We find that for most of the examples $\chi$ takes bigger values for singular entangling regions, which may suggest increase in distillable entanglement. However, there also exist counterexamples where distillable entanglement decreases for singular surfaces. We also explore the behaviour of $\chi$ as the coupling varies and observe that for singular entangling surfaces, the amount of distillable entanglement is mostly largest for free theories, while counterexample exists for free Dirac fermion in three dimensions. For holographic CFTs described by higher derivative gravity, $\chi$ may increase or decrease, depending on the sign of the relevant parameters. Our results may reveal a more profound connection between geometry and distillable entanglement.
\end{abstract}
\setcounter{page}{0}
\end{titlepage}

\pagestyle{plain} \baselineskip=19pt

\tableofcontents

\section{Introduction}
Entanglement may be the most mysterious phenomenon inherently related to quantum mechanics. Roughly speaking entanglement describes the presence of correlations, while the nature of entanglement can be either classical or quantum. So given a quantum state, a central question is to distinguish and quantify quantum entanglement from the classical counterpart. By far it has been realised that at least for pure states, the nature of entanglement can be completely characterised by the well-known Bell/Clauser-Horne-Shimony-Holt(CHSH) inequalities, but the case of mixed states is much less clear.

Let us focus on a pure state $\rho$ of a bipartite system, whose Hilbert spaces are denoted $\mathcal{H}_{L}$ and $\mathcal{H}_{R}$. If $\rho$ can be expressed as
\begin{equation}
\rho=\sum_{i}p_{i}\rho^{L}_{i}\otimes\rho^{R}_{i},~~\sum_{i}p_{i}=1,~~p_{i}\geq0,
\end{equation}
then $\rho$ is separable, otherwise it is entangled. Since separable states can be produced using only local operations and classical communications (LOCC), they are classically correlated. However, for mixed states the situation becomes much more complex, for example, the intuition that the only states that satisfy the Bell inequalities are the separable ones fails for mixed states. Several entanglement measures have been proposed because of the intricate nature of mixed state entanglement, among which a computable one is the entanglement negativity (EN)~\cite{Vidal:2002cme}.

For a bipartite system, the entanglement negativity is determined by the absolute values of the partial transposed density matrix (a detailed definition will be given in the next section). Such a concept can also be generalised in the framework of relativistic quantum field theories~\cite{Verch:2004vj, Calabrese:2012ew, Calabrese:2012nk}, where computations of entanglement negativity in 1+1-dimensional QFTs amount to evaluating twist operator correlations in~\cite{Calabrese:2012ew, Calabrese:2012nk}. Furthermore, progress in recent years has enabled us to `geometrise' the entanglement entropy (EE) in the context of gauge/gravity duality~\cite{Ryu:2006bv, Ryu:2006ef}. Such progress has opened up new windows towards  a deeper understanding on the connections between geometry and entanglement.

It is therefore natural to ask to which extent we can extract new properties of EN of general $d$-dimensional conformal field theories in the framework of holography. In~\cite{Rangamani:2014ywa} it was pointed out that for an entangling region $\mathcal{A}$, the EN in a pure state possesses the following properties:
\begin{itemize}
\item The leading divergent term scales as the area of the entangling surface $\partial\mathcal{A}$, which is analogous to the case of EE;
\item The sub-leading divergent terms have an identical structure to that in the EE for the reduced density matrix $\rho_{\mathcal{A}}$;
\item The value of the negativity generally takes a larger value than the corresponding EE, whose difference was conjectured to be in a geometric factor.
\end{itemize}
In particular, the authors of~\cite{Rangamani:2014ywa} defined the following quantity,
\begin{equation}
\label{chi}
\chi=\Big|\frac{C^{\rm univ}[\mathcal{E}_{N}]}{C^{\rm univ}[S_{EE}]}\Big|,
\end{equation}
where $C^{\rm univ}[\cdot\cdot]$ denotes the universal part of the entanglement negativity $\mathcal{E}_{N}$ and the EE $S_{EE}$. It was claimed in~\cite{Rangamani:2014ywa} that $\chi$ gives a precise measure of the entanglement negativity for the ground state in terms of the EE. In other words, the difference between the negativity and the EE may be encoded in $\chi$ and $\chi$ should just depend on the geometry of the entangling surface $\partial\mathcal{A}$. The values of $\chi$ for free theories and Einstein gravity with spherical entangling region were compared in~\cite{Rangamani:2014ywa}, where it was found that at least for the examples studied, $\chi$ is always bigger than 1 and takes  a smaller value at strong coupling, which seems to suggest a decrease in distillable entanglement in the strong coupling regime\footnote{Decrease in $\chi$ at strong coupling may be due to the reduction of total entanglement  or simply by the reduction of entanglement negativity. Here we assume the latter case.~\cite{Rangamani:2014ywa} }. Furthermore, $\chi$ can be smaller than 1 if the geometry and topology of the entangling surface are complicated enough~\cite{Perlmutter:2015vma}.

\begin{figure}
    \centering
    \includegraphics[width=0.4\textwidth]{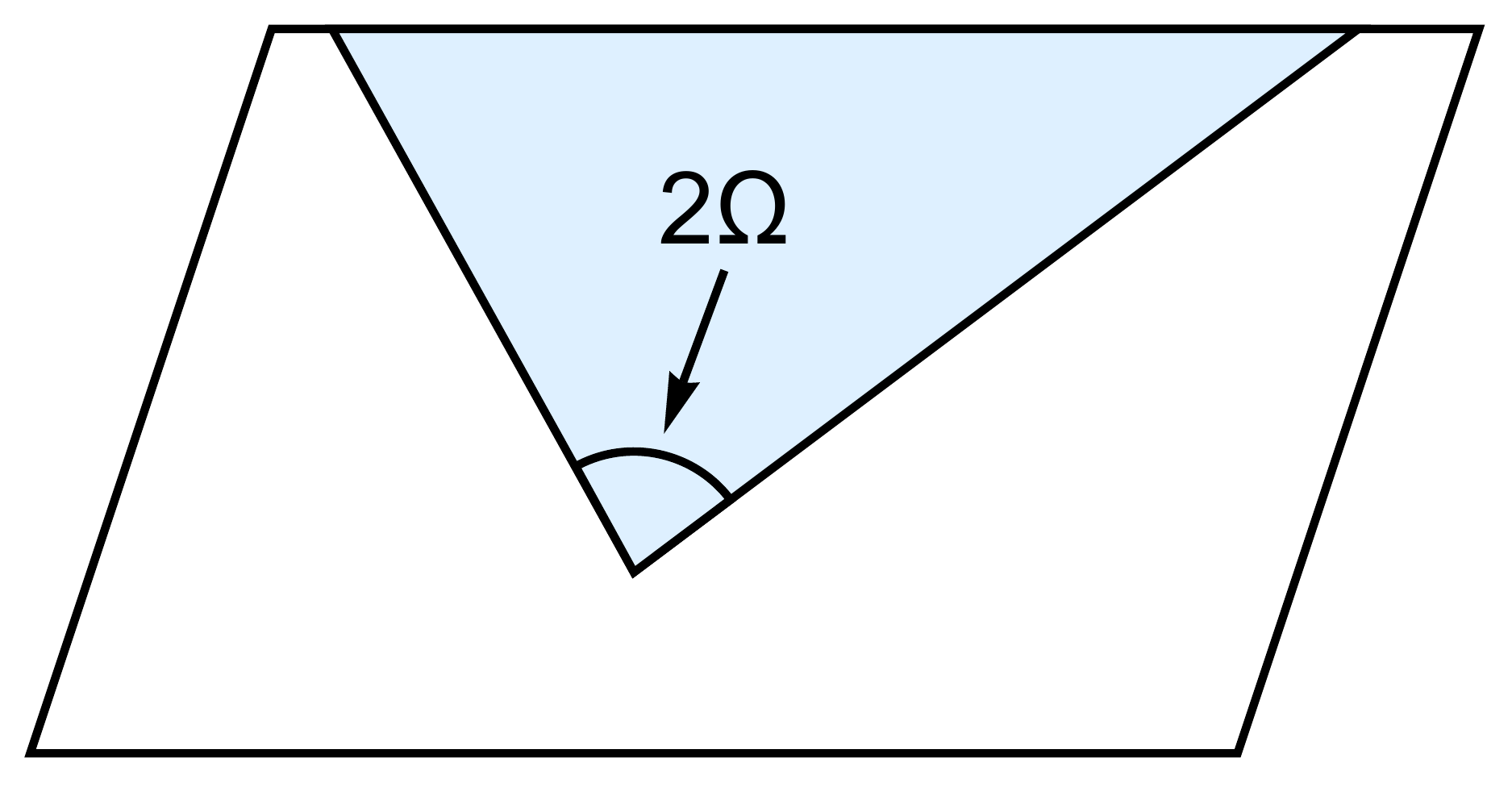}
    \caption{Entangling region that contains a corner.}
    \label{fig0}
\end{figure}

 In this paper, we further explore the dependence of $\chi$  on the geometry of entangling region. In particular, we extend \cite{Rangamani:2014ywa} to the case of entangling regions that contain corners as shown in Fig. \ref{fig0}. 
The main reason why entangling regions with corners are interesting is that they would lead to new universal contributions to the EE and R\'{e}nyi entropies, for example, for massless free scalars in 2+1-dimensions the corner-induced universal contributions read~\cite{Casini:2006hu}
\begin{equation}
S^{\rm univ}_{n}\sim a_{n}(\Omega)\log\left(\frac{H}{\epsilon}\right),
\end{equation}
where $H$ denotes the size of the entangling region, $\epsilon$ is the UV cutoff and $a_{n}(\Omega)$ is a function of the opening angle $\Omega$ of the corner. A similar structure for 2+1-dimensional free Dirac fermions was derived in~\cite{Casini:2008as}. Such a logarithmic term in the EE was also observed in the holographic setup, first in~\cite{Hirata:2006jx} in $AdS_{4}$ and subsequently in general $AdS_{d+1}$ in~\cite{Myers:2012vs}. Note that universal terms of the form $\log^{2}(H/\epsilon)$ may exist when $d$ is even.

Recently universal corner contributions to EE were revisited in~\cite{Bueno:2015rda, Bueno:2015xda}, where an elegant expression that relates the universal corner contribution to the EE $\sigma$ in 2+1-dimensions and the central charge of the corresponding CFT $C_{T}$ was derived
\begin{equation}
\label{sctsec1}
\sigma=\frac{\pi^{2}}{24}C_{T}.
\end{equation}
Here $\sigma$ is determined by the $\Omega$-dependent coefficient of the universal corner contribution in the vicinity of $\Omega=\pi/2$, i.e., in smooth limit, 
\begin{equation}
a_{1}(\Omega\rightarrow\pi/2)=4\sigma(\Omega-\pi/2)^{2}.
\end{equation}
The relation~(\ref{sctsec1}) was confirmed by free field examinations in~\cite{Elvang:2015jpa} and later by evaluating the entanglement entropy across a deformed planar or spherical entangling surface in terms of a perturbative expansion in the infinitesimal shape deformation in~\cite{Faulkner:2015csl, Bianchi:2015liz}. For other recent developments along this direction see~\cite{Pang:2015lka}-\cite{Bueno:2015ofa}.

As we will see, the presence of corners also leads to interesting results in the entanglement negativity, especially in~(\ref{chi}). We consider free complex scalar and free Dirac fermions in $d=3,4$, as well as strongly coupled CFTs evaluated via the corresponding gravity duals. In the holographic setup we study both Einstein gravity and higher derivative gravity: in $d=3$ the higher derivative terms contain squares of Ricci tensor and Ricci scalar and cubic in Weyl tensor; in $d=4$ we focus on quasi-topological gravity, which can be used to holographically describe CFTs with unequal central charges. We observe that for most of the examples, $\chi$ takes a bigger value when the entangling surface contains corners, which might indicate that singular entangling geometry increases the amount of distillable entanglement. However, there also exist counterexamples: free complex scalars in $d=3,4$ and for quasi-topological gravity in $d=4$ in a small parameter region. We also investigate the behaviour of $\chi$ as the coupling varies and find that for complex scalars in $d=3,4$ and Dirac fermions in $d=4$, $\chi$ still takes a larger value in the presence of corners, while for Dirac fermions in $d=3$ $\chi$ is slightly smaller than the strong coupling result, which might be taken as a counterexample of the statement proposed in~\cite{Rangamani:2014ywa}. Once higher derivative corrections are included, we find that the value of $\chi$ may increase or decrease, depending on the sign of the parameter in the theory.

The rest of the paper is organised as follows: In Section 2 we briefly review the definitions of various entanglement measures: entanglement entropy, R\'{e}nyi entropy and entanglement negativity,  and universal contributions to these measures. Then we study the corner contribution to entanglement measures $d=3$ CFTs in Section 3, beginning with analysis in free theories and followed by calculations in Einstein gravity and higher derivative gravity. A parallel investigation will be performed for $d=4$ CFTs in Section 4 and a summary and discussion will be presented in Section 5. 
\section{Universal contributions to entanglement measures}
In this section we will first briefly discuss the entanglement measures we are interested in: the entanglement entropy (EE), the R\'{e}nyi entropy (RE) and the entanglement negativity (EN), followed by a short introduction of the method for evaluating the RE developed in~\cite{Hung:2011nu}. Then we will collect new universal contributions to entanglement measures arising from singular entangling surfaces~\cite{Bueno:2015qya, Bueno:2015lza}.

The most widely investigated entanglement measure is the EE. For a subsystem $A$ equipped with the reduced density matrix $\rho_{A}$, the EE is defined by,
\begin{equation}
S_{\rm EE}=-{\rm Tr}(\rho_{A}\log\rho_{A}).
\end{equation}
Another interesting entanglement measure is the RE, which is given by
\begin{equation}
\label{renyisec2}
S_{n}=\frac{1}{1-n}{\rm Tr}\log\rho_{A}^{n}.
\end{equation}
The RE can lead to many novel results when $n$ takes different values, for instance, $\lim_{n\rightarrow1}S_{n}=S_{\rm EE}$.

Comparing with the EE and the RE, the EN (denoted by $\mathcal{E}_{N}$) has not been well understood. The EN can be used to define a measure of the amount of distillable entanglement in a particular state.
Consider a bipartite system whose total Hilbert space is a tensor product of the left- and right-ones $\mathcal{H}_{L}\otimes\mathcal{H}_{R}$, where the basis of each sub-Hilbert space are given by $|{\bf r}_{a}\rangle$
and $|{\bf l}_{\alpha}\rangle$ respectively, with $a\in\{1,2,\cdots,{\rm dim}(\mathcal{H}_{R})\}$ and $\alpha\in\{1,2,\cdots,{\rm dim}(\mathcal{H}_{L})\}$. In this basis the matrix elements of a general density matrix $\rho$ can be expressed as
\begin{equation}
\rho_{a\alpha,b\beta}=\langle{\bf r}_{a}{\bf l}_{\alpha}|\rho|{\bf r}_{b}{\bf l}_{\beta}\rangle.
\end{equation}
To define the EN, we take the partial transpose with respect to the left subsystem without loss of generality and obtain the following partial transposed density matrix $\rho^{\Gamma}$,
\begin{equation}
\rho^{\Gamma}_{a\alpha,b\beta}=\rho_{a\beta,b\alpha}=\langle{\bf r}_{a}{\bf l}_{\beta}|\rho|{\bf r}_{b}{\bf l}_{\alpha}\rangle,
\end{equation}
and the negativity $\mathcal{N}(\rho)$ and the logarithmic negativity $\mathcal{E}(\rho)$ are given by
\begin{equation}
\mathcal{N}(\rho)=\frac{||\rho^{\Gamma}||_{1}-1}{2},~~\mathcal{E}_{N}=\log||\rho^{\Gamma}||_{1},
\end{equation}
where for an operator $\mathcal{O}$
\begin{equation}
\label{osq}
||\mathcal{O}||_{1}={\rm Tr}\left(\sqrt{\mathcal{O}^{\dag}\mathcal{O}}\right).
\end{equation}
The negativity $\mathcal{E}(\rho)$ measures the number of negative eigenvalues of $\rho^{\Gamma}$.

Generically it is very daunting to evaluate the EN because of the square root in the definition~(\ref{osq}). However, for any pure state $\psi=|\Psi\rangle\langle\Psi|$ of a bipartite system the EN can be expressed in a very simple form~\cite{Calabrese:2012nk}
\begin{equation}
\mathcal{E}_{N}(\psi)=S_{1/2}(\rho^{R,L}),
\end{equation}
where $\rho^{L,R}$ denotes the reduced density matrix associated with the left/right subsystem $\rho^{L,R}={\rm Tr}_{R,L}(\rho)$. In other words, in such a system the EN is given by the RE with index $1/2$, which can be evaluated via the method in~\cite{Casini:2011kv}: for a spherical entangling region $A$ with radius $R$, centered without loss of generality at the origin of $d$-dimensional Minkowski spacetime
\begin{equation}
ds^{2}=-dt^2+dr^{2}+r^{2}d\Omega_{d-2}^{2},
\end{equation}
it can be shown that the reduced density matrix $\rho_{A}$ is equivalent to the thermal density matrix for the CFT on $R\times H^{d-1}$ at $T_{0}=1/(2\pi R)$ upon conformal mapping. Furthermore, this method can also be generalised to calculations of RE~\cite{Hung:2011nu}
\begin{equation}
\label{resec2}
 S_{n}=\frac{n}{1-n}\frac{1}{T_{0}}[F(T_{0})-F(T_{0}/n)],
\end{equation}
where $F(T)$ is the corresponding free energy at temperature $T$.

One reason that universal contributions to entanglement measures are of great interest is: the underlying degrees of freedom of the system in question may be encoded in these universal contributions. For smooth entangling geometries in a $d$-dimensional QFT, the universal contributions to entanglement measures can be schematically expressed as
\begin{eqnarray}
E&=&\sum\limits^{d-4}_{k=0}\frac{E_{k}}{\epsilon^{d-2-2k}}+(-1)^{\frac{d-1}{2}}C^{\rm univ}[E],~~d=~{\rm odd}\nonumber\\
E&=&\sum\limits^{d-4}_{k=0}\frac{E_{k}}{\epsilon^{d-2-2k}}+(-1)^{\frac{d-2}{2}}C^{\rm univ}[E]\log\frac{H}{\epsilon}+C_{0},~~d=~{\rm even}
\end{eqnarray}
where $E$ denotes $\{S_{\rm EE}, S_{n}, \mathcal{E}_{N}\}$ collectively, $H$ is the size of the system and $\epsilon$ is the UV cut-off.
In~\cite{Rangamani:2014ywa} it was observed that the following quantity constructed from the universal contributions to the EN and the EE
\begin{equation}
\chi=\Big|\frac{C^{\rm univ}[\mathcal{E}_{N}]}{C^{\rm univ}[S_{EE}]}\Big|
\end{equation}
exhibits certain monotonicity, that is, it becomes smaller as the coupling constant increases. Supporting evidences include results from free complex scalars and fermions in $d=3$ and $\mathcal{N}=4$ SYM in $d=4$, as well as holographic CFTs described in terms of Einstein gravity. It was argued that $\chi$ should depend just on the geometry of the entangling surface~\cite{Rangamani:2014ywa}

To further understand the dependence of $\chi$ on geometry, here we consider a simpler class of entangling regions: surfaces that contain corners. Recently it was shown in~\cite{Bueno:2015qya, Bueno:2015lza} that corners can induce new universal contributions to the RE
\begin{eqnarray}
S^{\rm univ}_{n}&=&(-1)^{\frac{d-1}{2}}a^{(d)}_{n}(\Omega)\log\frac{H}{\epsilon},~~d=~{\rm odd}\nonumber\\
S^{\rm univ}_{n}&=&(-1)^{\frac{d-2}{2}}a^{(d)}_{n}(\Omega)\log^{2}\frac{H}{\epsilon},~~d=~{\rm even},
\end{eqnarray}
where $\Omega$ is the opening angle $\Omega\in[0,\pi]$ and the coefficient $a_{n}^{(d)}$ satisfies $a_{n}^{(d)}(\Omega)=a_{n}^{(d)}(\pi-\Omega)$. Moreover, if we take the smooth limit, i.e., in the vicinity of $\Omega=\pi/2$, we can obtain
\begin{equation}
a^{(d)}_{n}(\Omega\rightarrow\pi/2)=4\sigma^{(d)}_{n}(\Omega-\pi/2)^{2}.
\end{equation}
In other words, the universal corner contribution defines a set of coefficients $\sigma^{(d)}_{n}$ which encode regulator-independent information about the underlying QFT. It was further conjectured in~\cite{Bueno:2015lza}
that
\begin{equation}
\label{sdn}
\sigma^{(d)}_{n}=g(d)\frac{h_{n}}{n-1},
\end{equation}
where $h_{n}$ is the scaling dimension of the twist operator~\cite{Hung:2011nu}
\begin{equation}
\label{hnsec2}
h_{n}=\frac{2\pi nR^{d}}{d-1}(\mathcal{E}(T_{0})-\mathcal{E}(T_{0}/n)).
\end{equation}
In~(\ref{hnsec2})~$\mathcal{E}=E/(R^{d-1}V_{H})$ denotes the energy density ($E$ is the energy) and $V_{H}$ denotes the volume of the hyperboloid
\begin{eqnarray}
V_{H}&=&\Omega_{d-2}\int^{\frac{R}{\epsilon}}_{1}(y^{2}-1)^{(d-3)/2}dy\nonumber\\
&\simeq&\frac{\Omega_{d-2}}{d-2}\left[\frac{R^{d-2}}{\epsilon^{d-2}}-\frac{(d-2)(d-3)}{2(d-4)}\frac{R^{d-4}}{\epsilon^{d-4}}+\cdots\right].
\end{eqnarray}
$g(d)$ in~(\ref{sdn}) is a numerical factor, whose explicit expressions are shown as follows for completeness
\begin{eqnarray}
g(d={\rm even})&=&\frac{(d-1)(d-2)\pi^{\frac{d-4}{2}}\Gamma(\frac{d-1}{2})^{2}}{16\Gamma(\frac{d}{2})^{3}},\nonumber\\
g(d={\rm odd})&=&g(d={\rm even})\times\pi.
\end{eqnarray}
Since corners can lead to new universal contributions to RE, how do they affect the behaviour of $\chi$? In particular, will the monotonicity still hold? We will try to answer such questions in the following sections.

\section{Corner contributions in 3d CFTs}
In this section we evaluate the new contributions to $\chi$ induced by the corner in 3d CFTs, where $\chi$ has been defined in~(\ref{chi}). We also compare those results with the counterparts for smooth entangling surfaces, aiming at understanding the effect of corners on distillable entanglement. Firstly we consider free complex bosons and fermions in subsection 3.1, then in subsection 3.2 we compare the free results with those obtained from Einstein gravity. Finally in subsection 3.3 we study the higher derivative corrections to $\chi$ by considering general quadratic and cubic terms in gravity.

Before proceeding we give a brief summary of our notations: In the superscripts $(d)$ denotes the number of spacetime dimensions on the field theory side, `cs/f' stands for complex scalar/fermion and `E/H' denotes holographic results evaluated in Einstein/higher derivative gravity; in the subscripts `sm' denotes results for smooth entangling geometries while `sig' stands for those for singular entangling surfaces. For simplicity we will just include the cut-off independent terms in the results for the EEs and REs from now on.
\subsection{Free theory}
Let us first look at free theory. Using~(\ref{resec2}) the R\'{e}nyi entropy can be rewritten as follows
\begin{equation} \label{new1}
S_{n}=\frac{n\mathcal{F}_{1}-\mathcal{F}_{n}}{1-n},~~\mathcal{F}_{n}\equiv2\pi RnF(T_{0}/n), 
\end{equation}
which leads to
\begin{equation}
S_{1/2}=\mathcal{F}_{1}-2\mathcal{F}_{1/2},~~S_{1}=-\mathcal{F}_{1}.
\end{equation}
For free complex scalar we have closed form for $\mathcal{F}$~\cite{Klebanov:2011uf}
\begin{equation}
\mathcal{F}^{(3),{\rm cs}}_{n}=-\int^{\infty}_{0}d\lambda\tanh(\pi\sqrt{\lambda})\log(1-e^{-2\pi n\sqrt{\lambda}})+n\frac{3\zeta(3)}{4\pi^{2}},
\end{equation}
from which we can easily read off the results needed
\begin{equation}
\mathcal{F}^{(3),{\rm cs}}_{1/2}\approx0.237094,~~~\mathcal{F}^{(3),{\rm cs}}_{1}\approx0.127614.
\end{equation}
Hence the REs are given by
\begin{equation}
S^{(3),{\rm cs}}_{1/2}\approx-0.346574,~~S^{(3),{\rm cs}}_{1}\approx-0.127614.
\end{equation}
Therefore for smooth entangling region we have
\begin{equation}
\chi^{(3),{\rm cs}}_{\rm sm}=\frac{S^{(3),{\rm cs}}_{1/2}}{S^{(3),{\rm cs}}_{1}}=2.71579.
\end{equation}

The case of free fermions can be analysed in a parallel way, from which we have~\cite{Klebanov:2011uf}
\begin{equation}
\mathcal{F}^{(3),{\rm f}}_{n}=2\int^{\infty}_{0}d\lambda\lambda\coth(\pi\lambda)\log(1+e^{-2\pi n\lambda})+n\frac{\zeta(3)}{\pi^{2}},
\end{equation}
which enables us to extract
\begin{equation}
\mathcal{F}^{(3),{\rm f}}_{1/2}\approx0.316125,~~\mathcal{F}^{(3),{\rm f}}_{1}\approx0.218959.
\end{equation}
Therefore the REs for free fermions read
\begin{equation}
S^{(3),{\rm f}}_{1/2}\approx-0.413291,~~S^{(3),{\rm f}}_{1}\approx-0.218959,
\end{equation}
and the ratio $\chi$ is given by
\begin{equation}
\chi^{(3),{\rm f}}_{\rm sm}=\frac{S^{(3),{\rm f}}_{1/2}}{S^{(3),{\rm f}}_{1}}=1.88752.
\end{equation}

Generally closed forms for the coefficient of universal corner contribution $\sigma_{n}$ can only be obtained in certain specific cases. In~\cite{Elvang:2015jpa} it was observed that for complex scalar and Dirac fermion in $d=3$ at integer $n$, we have
\begin{equation}
\sigma^{(3),{\rm cs}}_{n}=\sum\limits^{n-1}_{k=1}\frac{k(n-k)(n-2k)\tan\frac{\pi k}{n}}{12\pi n^{3}(n-1)},~~\sigma^{(3),{\rm f}}_{n}=\sum\limits^{(n-1)/2}_{k=-(n-1)/2}\frac{k(n^{2}-4k^{2})\tan\frac{\pi k}{n}}{24\pi n^{3}(n-1)}.
\end{equation}
In particular, the universal corner contribution to the EE reads
\begin{equation}
\sigma^{(3),{\rm cs}}_{1}=\sigma^{(3),{\rm f}}_{1}=\frac{1}{128}.
\end{equation}
To obtain the REs with index $1/2$, we may apply the `Bose-Fermi duality' between $\sigma_{n}$s~\cite{Bueno:2015qya}
\begin{equation}
n^{2}\sigma^{(3),{\rm cs/f}}_{n}=\sigma^{(3),{\rm f/cs}}_{1/n},
\end{equation}
which leads to
\begin{equation}
\sigma^{(3),{\rm cs}}_{1/2}=4\sigma^{(3),{\rm f}}_{2}=\frac{1}{16\pi}, \qquad \sigma^{(3),{\rm f}}_{1/2}=4\sigma^{(3),{\rm cs}}_{2}=\frac{1}{6\pi^{2}}.
\end{equation}
Therefore, for singular entangling surfaces we have
\begin{equation}
\chi^{(3),{\rm cs}}_{\rm sig} = {\frac{\sigma^{(3),\rm cs}_{1/2}}{\sigma^{(3),\rm cs}_{1}}}  = \frac{8}{\pi}\approx2.54648, \qquad \chi^{(3),{\rm f}}_{\rm sig}=  {\frac{\sigma^{(3),\rm f}_{1/2}}{\sigma^{(3),\rm f}_{1}}}  = \frac{64}{3\pi^{2}}\approx2.16152.
\end{equation}

\subsection{Holographic considerations: Einstein gravity}
To explore the behaviour of $\chi$ in the strong coupling limit, in this subsection we consider the results obtained via holography in the large $N$ limit, whose dual description is given by Einstein gravity.
In fact the main results were obtained in~\cite{Hung:2011nu} and here we collect them and compare them with the free theory counterparts.

According to~\cite{Casini:2011kv, Hung:2011nu} the RE of a CFT on a sphere can be related to the entropy of topological AdS black holes. For our interest the topological AdS black hole metric is given by
\begin{equation}
ds^{2}=-\left(\frac{r^{2}}{L^{2}}f(r)-1\right)N(r)^{2}dt^{2}+\frac{dr^{2}}{\frac{r^{2}}{L^{2}}f(r)-1}+r^{2}d\Sigma_{2}^{2},
\end{equation}
where
\begin{equation}
f(r)=1-\frac{\omega^{3}}{r^{3}},~~N(r)=\frac{L}{R},
\end{equation}
and $d\Sigma_{2}^{2}$ denotes the line element on a hyperbolic space $H^{2}$. The expression for the REs of a $d$-dimensional CFT reads
\begin{equation}
\label{renyiein}
S^{(d),{\rm E}}_{n}=\frac{L^{d-1}}{8G}V_{H}\frac{n}{n-1}\left(2-x_{n}^{d-2}(1+x_{n}^{2})\right),
\end{equation}
where
\begin{equation}
\label{xnein}
x_{n}=\frac{1}{nd}\left(1+\sqrt{1-2dn^{2}+d^{2}n^{2}}\right).
\end{equation}
Furthermore, using~(\ref{hnsec2}) we can also obtain the scaling dimension
\begin{equation}
\label{hnein}
h_{n}=\frac{L^{d-1}}{8G}nx_{n}^{d-2}(1-x_{n}^{2}).
\end{equation}

All the above results enable us to obtain $\chi$ for both smooth and singular entangling surfaces
\begin{equation}
\chi^{(3),{\rm E}}_{\rm sm}=\frac{S^{(3),{\rm E}}_{1/2}}{S^{(3),{\rm E}}_{1}}\approx1.63113,
\end{equation}
\begin{equation}
\chi^{(3),{\rm E}}_{\rm sig}=\frac{\sigma^{(3),{\rm E}}_{1/2}}{\sigma^{(3),{\rm E}}_{1}} = {-2\frac{h_{1/2}^{(3), \rm E}}{\partial_n h^{(3), \rm E}_n |_{n \rightarrow 1}}} \approx2.16509.
\end{equation}
Combining the holographic results and the free field results obtained in the last subsection, we can see
\begin{itemize}
\item For free theories we have
\begin{equation}
\chi^{(3),{\rm cs}}_{\rm sm}>\chi^{(3),{\rm cs}}_{\rm sig},~~\chi^{(3),{\rm f}}_{\rm sig}>\chi^{(3),{\rm f}}_{\rm sm},
\end{equation}
which suggests that for complex scalars, the presence of corners reduces the amount of distillable entanglement while for fermions the situation reverses.
\item For holographic theories we have
\begin{equation}
\chi^{(3),{\rm E}}_{\rm sig}>\chi^{(3),{\rm E}}_{\rm sm},
\end{equation}
which may indicate that in the strong coupling limit, the amount of distillable entanglement is increased by the corners.
\item For complex scalars in different regimes of coupling,
\begin{equation}
  \chi^{(3),{\rm cs}}_{\rm sm}>\chi^{(3),{\rm E}}_{\rm sm},~~\chi^{(3),{\rm cs}}_{\rm sig}>\chi^{(3),{\rm E}}_{\rm sig},
\end{equation}
which means that the monotonicity of $\chi$ (decreases as the coupling increases) is preserved even for singular entangling surfaces.
\item The most curious phenomenon exhibits for free fermions,
\begin{equation}
\chi^{(3),{\rm f}}_{\rm sm}>\chi^{(3),{\rm E}}_{\rm sm},~~\chi^{(3),{\rm f}}_{\rm sig}<\chi^{(3),{\rm E}}_{\rm sig},
\end{equation}
which tells us that for smooth entangling regions the monotonicity still holds, as observed in~\cite{Rangamani:2014ywa}, but for singular entangling regions the monotonicity
is slightly violated, indicating a small increase in the amount of distillable entanglement as the coupling becomes stronger.
\end{itemize}
\subsection{Holographic considerations: Higher derivative gravity}
Generally both $1/N$ and finite-coupling corrections on the field theory side would induce higher derivative corrections on the dual gravity side, which will result in interesting results. In this subsection we consider finite $N$/finite-coupling effects to $\chi$ by considering higher derivative gravity to see whether the monotonicity can be violated by higher derivative corrections. It should be pointed out that since we are considering bottom-up models, whose field theory duals are not completely known, we cannot distinguish finite $N$ or finite-coupling corrections in the present setup. A general $d+1$-dimensional action containing quadratic and cubic in curvature can be written as \cite{Myers:2010ru}.
\begin{equation}
\label{qtg}
I=\frac{1}{16\pi G}\int d^{d+1}x\sqrt{-g}\left[R+\frac{d(d-1)}{L^2}+L^{2}\tilde{\chi}+L^{4}\tilde{\mathcal{Z}}\right],
\end{equation}
where
\begin{equation}
\tilde{\chi}=\lambda_{1}R_{abcd}R^{abcd}+\lambda_{2}R_{ab}R^{ab}+\lambda_{3}R^{2},
\end{equation}
\begin{eqnarray}
\tilde{\mathcal{Z}}&=&\mu_{1}{{{R_{a}}^{c}}_{b}}^{d}{{{R_{c}}^{e}}_{d}}^{f}{{{R_{e}}^{a}}_{f}}^{b}+\mu_{2}{R_{ab}}^{cd}{R_{cd}}^{ef}{R_{ef}}^{ab}+\mu_{3}R_{abcd}{R^{abc}}_{e}R^{de}\nonumber\\
& &+\mu_{4}R_{abcd}R^{abcd}R+\mu_{5}R_{abcd}R^{ac}R^{bd}+\mu_{6}{R_{a}}^{b}{R_{b}}^{c}{R_{c}}^{a}\nonumber\\
& &+\mu_{7}{R_{a}}^{b}{R_{b}}^{a}R+\mu_{8}R^{3}.
\end{eqnarray}
In addition, there can be other candidates constructed from the Weyl tensor
\begin{equation}
\mathcal{W}_{1}={{{C_{a}}^{c}}_{b}}^{d}{{{C_{c}}^{e}}_{d}}^{f}{{{C_{e}}^{a}}_{f}}^{b},~~\mathcal{W}_{2}={C_{ab}}^{cd}{C_{cd}}^{ef}{C_{ef}}^{ab}.
\end{equation}
It was pointed out in~\cite{Sinha:2010pm, Myers:2010tj} that $\tilde{\mathcal{Z}}$ is not well-defined in $d=3$ and $\mathcal{W}_{1}=\mathcal{W}_{2}$ as implied by Schouten identities,  so we will consider the following action
\begin{equation}
\label{hd3d}
I=\frac{1}{16\pi G}\int d^{4}x\sqrt{-g}\left[R+\frac{6}{L^{2}}+L^{2}(\lambda_{1}R^{2}+\lambda_{2}R_{ab}R^{ab})+\lambda_{3}L^{3}{C_{ab}}^{cd}{C_{cd}}^{ef}{C_{ef}}^{ab}\right].
\end{equation}
One may also add a Riemann curvature squared term $\lambda_{4}R_{abcd}R^{abcd}$ in~(\ref{hd3d}). However, since the Gauss-Bonnet invariant $E_{4}=R_{abcd}R^{abcd}-4R_{ab}R^{ab}+R^{2}$ does not contribute to the equations of motion in $d=3$, one can eliminate the Riemann squared term by taking $E_{4}$ into account, leaving the general action~({\ref{hd3d}}).

Given the ansatz for the metric
\begin{eqnarray}
ds^{2}&=&-\left(\frac{r^{2}}{L^{2}}f(r)-1\right)N(r)^{2}dt^{2}+\frac{dr^{2}}{\frac{r^{2}}{L^{2}}f(r)-1}+r^{2}d\Sigma_{2}^{2},\nonumber\\
f(r)&=&1-\frac{\omega^{3}}{r^{3}}-\sum\limits^{3}_{i=1}\lambda_{i}f_{i}(r),~~N(r)=\frac{L}{R}\left(1-\sum\limits_{i=1}^{3}\lambda_{i}N_{i}(r)\right),
\end{eqnarray}
we can work out the perturbative black hole solution
\begin{eqnarray}
& &f_{1}(r)=f_{2}(r)=0,~~f_{3}(r)=\frac{2\omega^{6}}{r^{9}}(8\omega^{3}+9L^{2}r-12r^{3}),\nonumber\\
& &N_{1}(r)=N_{2}(r)=0,~~N_{3}(r)=\frac{6\omega^{6}}{r^{6}}.
\end{eqnarray}
The black hole horizon is located at $r=r_{H}$ such that $f(r_{H})=L^{2}/r_{H}^{2}$. For convenience we express $\omega$ in terms of $r_{H}$ to first order in $\lambda_{i}$,
\begin{equation}
\omega^{3}=r_{H}^{3}-L^{2}r_{H}+\lambda_{3}\frac{2(4r_{H}^2-L^{2})(r_{H}^{2}-L^{2})^{2}}{r_{H}^{3}}.
\end{equation}
The black hole temperature is given by
\begin{eqnarray} \label{TTT}
T&=&\frac{N(r_{H})}{4\pi}\left[\frac{2}{r_{H}}+\frac{r_{H}^{2}}{L^{2}}\frac{\partial f(r)}{\partial r}\Big|_{r=r_{H}}\right]\nonumber\\
&=&\frac{1}{4\pi R}\left(3x-\frac{1}{x}+6\lambda_{3}\frac{(1-x^{2})^{2}}{x^{3}}\right),
\end{eqnarray}
where $x\equiv r_{H}/L$. The black hole entropy is given by Wald formula
\begin{equation}
S_{\rm BH}=-2\pi\int d^{d-1}y\sqrt{h}\frac{\partial\mathcal{L}}{\partial R_{mnpq}}\epsilon_{mn}\epsilon_{pq}  \,,
\end{equation}
with $\epsilon$ the unit binormal vector, which leads to
\begin{equation}
S_{\rm BH}=\frac{L^{2}V_{H}}{4G}\left[(1-24\lambda_{1}-6\lambda_{2})x^{2}+6\lambda_{3}\frac{(1-x^{2})^{2}}{x^{2}}\right].
\end{equation}
Hence we can evaluate the RE via~\cite{Hung:2011nu}
\begin{equation}
S_{n}=\frac{2\pi Rn}{n-1}\int^{1}_{x_{n}}S_{\rm BH}(x)\frac{dT(x)}{dx}dx,
\end{equation}
where $x_{n}$ is determined by $T(x_{n})=1/(2\pi Rn)$. 
{To first order in $\lambda_{i}$}, we have
\begin{equation} \label{xn3}
x_{n}=\frac{1+\sqrt{1+3n^{2}}}{3n}-\frac{4\lambda_{3}(\sqrt{1+3n^{2}}-2)^{2}}{3n\sqrt{1+3n^{2}}}.
\end{equation}
Therefore, the REs in the presence of higher derivative terms read
\begin{equation}
S^{(3),{\rm H}}_{n}=\frac{L^{2}V_{H}}{8G}\frac{n}{n-1}\left[(2-x_{n}-x_{n}^{3})(1-24\lambda_{1}-6\lambda_{2})+2\lambda_{3}\frac{(1-4x_{n}^{2})(1-x_{n}^{2})^{2}}{x_{n}^{3}}\right],
\end{equation}
and we can obtain the explicit expressions for the REs in different limits
\begin{eqnarray}
\lim_{n\rightarrow0}S^{(3),{\rm H}}_{n}&=&\frac{L^{2}V_{H}}{27G}(1-24\lambda_{1}-6\lambda_{2}-4\lambda_{3}),\\
\lim_{n\rightarrow1}S^{(3),{\rm H}}_{n}&=&\frac{L^{2}V_{H}}{4G}(1-24\lambda_{1}-6\lambda_{2}),\\
\lim_{n\rightarrow\infty}S^{(3),{\rm H}}_{n}&=&\frac{L^{2}V_{H}}{36G}\left[(9-2\sqrt{3})(1-24\lambda_{1}-6\lambda_{2})+8\sqrt{3}\lambda_{3}\right].
\end{eqnarray}
Therefore, for smooth entangling regions
\begin{equation} \label{xx}
\chi^{(3),{\rm H}}_{\rm sm}=\frac{S^{(3),{\rm H}}_{1/2}}{S^{(3),{\rm H}}_{1}}\approx1.63113-3.05088\lambda_{3}.
\end{equation}

For singular entangling regions $\chi$ can be obtained through scaling dimension of twist operators~\cite{Hung:2011nu}
\begin{equation}
h_{n}=\frac{2\pi Rn}{d-1}\int^{1}_{x_{n}}T(x)\frac{dS_{BH}(x)}{dx}dx,
\end{equation}
which leads to
\begin{equation}
h^{(3),{\rm H}}_{n}=\frac{L^{2}}{8G}n\left[x_{n}(1-x_{n}^{2})(1-24\lambda_{1}-6\lambda_{2})+2\lambda_{3}\frac{(1-x_{n}^{2})(1-4x_{n}^{2})}{x_{n}^{3}}\right],
\end{equation}
and
\begin{equation}
\lim_{n\rightarrow1}\partial_{n}h^{(3),{\rm H}}_{n}=\frac{L^{2}}{8G}(1-24\lambda_{1}-6\lambda_{2}).
\end{equation}
Finally it can be seen that
\begin{equation}
\chi^{(3),{\rm H}}_{\rm sig}=\frac{\sigma^{(3),{\rm H}}_{1/2}}{\sigma^{(3),{\rm H}}_{1}}=  {-2\frac{h_{1/2}^{(3), \rm H}}{\partial_n h^{(3), \rm H}_n |_{n \rightarrow 1}}} = 2.16509-2.40478\lambda_{3}.
\end{equation}

From those results we may conclude that at least in our examples, even though the REs and scaling dimensions depend on all $\lambda_{i}$, $\chi$ only depends on $\lambda_{3}$.  Moreover,
depending on the sign of $\lambda_{3}$, both $\chi^{(3),{\rm H}}_{\rm sm}$ and $\chi^{(3),{\rm H}}_{\rm sig}$ can be either larger or smaller than the counterparts in Einstein gravity, which means that finite $N$/finite-coupling effects may either increase~($\lambda_{3}<0$) or decrease~($\lambda_{3}>0$) the amount of distillable entanglement. In other words, the monotonicity may be broken by positive $\lambda_{3}$. We may not fix the allowed range of $\lambda_{3}$ by physical constraints such as causality, as we are working in perturbative expansion.
To check if singular entangling surface would increase the amount of distillable entanglement we compute the difference between $\chi^{(3),{\rm H}}_{\rm sm}$ and $\chi^{(3),{\rm H}}_{\rm sig}$:
\begin{equation}
\chi^{(3),{\rm H}}_{\rm sm}-\chi^{(3),{\rm H}}_{\rm sig}\approx-0.53396-0.6461\lambda_{3}.
\end{equation}
For $\lambda_3>-0.826436$, $\chi^{(3),{\rm H}}_{\rm sm}<\chi^{(3),{\rm H}}_{\rm sig}$ while for $\lambda_3<-0.826436$, $\chi^{(3),{\rm H}}_{\rm sm}>\chi^{(3),{\rm H}}_{\rm sig}$.
\section{Corner contributions in 4d CFTs}
In this section we evaluate corner contributions to $\chi$ in 4d CFTs, both in the free theory and the holographic Einstein gravity dual. The analysis is similar to what we did in Section 3 while the results are different, in particular, the monotonicity is preserved for both complex scalars and fermions. We also study the finite $N$/finite-coupling effect by considering quasi-topological gravity~\cite{Myers:2010ru, Myers:2010jv} and our numerical observations suggest that the monotonicity is still preserved when higher derivative terms are incorporated.
\subsection{Free theory}
Let us first consider complex scalar again. The `modified' free energy $\mathcal{F}$ reads~\cite{Klebanov:2011uf}
\begin{equation}
\label{fn4}
\mathcal{F}^{(4),{\rm cs}}_{n}=\frac{V_{H}}{2\pi^{2}}\int^{\infty}_{0}d\lambda\lambda^{2}\log(1-e^{-2\pi n\lambda}),
\end{equation}
where we have subtracted the flat space contribution to render $\mathcal{F}_{n}$ finite. Then we arrive at the REs by
substituting~(\ref{fn4}) into~(\ref{new1}),
\begin{equation}
S^{(4),{\rm cs}}_{n}=-\frac{(1+n)(1+n^{2})}{360n^{3}}\log\frac{H}{\epsilon},
\end{equation}
and hence $\chi$ for smooth entangling surfaces
\begin{equation}
\chi^{(4),{\rm cs}}_{\rm sm}=\frac{S^{(4),{\rm cs}}_{1/2}}{S^{(4),{\rm cs}}_{1}}=3.75.
\end{equation}
The universal corner contribution is determined by the scaling dimension of twist operators, which is given by~(\ref{hnsec2})
\begin{eqnarray}
\sigma^{(4),{\rm cs}}_{n}&=&\frac{g(4)n}{3(n-1)V_{H}}(\partial_{n}\mathcal{F}^{(4),{\rm cs}}_{n}|_{n=1}-\partial_{n}\mathcal{F}^{(4),{\rm cs}}_{n})\nonumber\\
&=&\frac{g(4)}{720\pi n^{3}}(1+n)(1+n^{2}).
\end{eqnarray}
Therefore one can obtain $\chi$ for singular surfaces
\begin{equation}
\chi^{(4),{\rm cs}}_{\rm sig}=\frac{\sigma^{(4),{\rm cs}}_{1/2}}{\sigma^{(4),{\rm cs}}_{1}}=3.75.
\end{equation}
This result can be verified by using the scaling dimension of twist operators obtained via heat kernel method in~\cite{Hung:2014npa}.

The case of free fermions can be analysed in a parallel way, which gives
\begin{equation}
\mathcal{F}^{(4),{\rm f}}_{n}=-\frac{V_{H}}{\pi^{2}}\int^{\infty}_{0}d\lambda(\lambda^{2}+\frac{1}{4})\log(1+e^{-2\pi n\lambda}).
\end{equation}
We can further obtain the RE for smooth entangling regions
\begin{equation}
S^{(4),{\rm f}}_{n}=-\frac{(1+n)(7+37n^{2})}{1440n^{3}}\log\frac{H}{\epsilon},
\end{equation}
as well as the universal corner contribution
\begin{equation}
\sigma^{(4),{\rm f}}_{n}=\frac{g(4)}{2880\pi}\frac{(1+n)(7+17n^{2})}{n^{3}}.
\end{equation}
Therefore $\chi$ for smooth and singular entangling regions are
\begin{equation}
\chi^{(4),{\rm f}}_{\rm sm}=\frac{S^{(4),{\rm f}}_{1/2}}{S^{(4),{\rm f}}_{1}}\approx2.21591,
\end{equation}
\begin{equation}
\chi^{(4),{\rm f}}_{\rm sig}=\frac{\sigma^{(4),{\rm f}}_{1/2}}{\sigma^{(4),{\rm f}}_{1}}=2.8125.
\end{equation}
\subsection{Holographic considerations: Einstein gravity}
Holographic calculations in Einstein gravity are straightforward: Substituting $d=4$ into~(\ref{renyiein}) and~(\ref{hnein}), we can obtain
\begin{equation}
\chi^{(4),{\rm E}}_{\rm sm}=\frac{S^{(4),{\rm E}}_{1/2}}{S^{(4),{\rm E}}_{1}}\approx1.67404,
\end{equation}
\begin{equation}
\chi^{(4),{\rm E}}_{\rm sig}=\frac{\sigma^{(4),{\rm E}}_{1/2}}{\sigma^{(4),{\rm E}}_{1}}\approx2.42404.
\end{equation}
Comparing the holographic results with the free field results in the previous subsection, we observe
\begin{itemize}
\item For free theories we have
\begin{equation}
\chi^{(4),{\rm cs}}_{\rm sm}=\chi^{(4),{\rm cs}}_{\rm sig},~~\chi^{(4),{\rm f}}_{\rm sig}>\chi^{(4),{\rm f}}_{\rm sm},
\end{equation}
which suggests that for complex scalars, the presence of corners does not affect the amount of distillable entanglement, which is quite different from the $d=3$ case. On the other hand, for fermions the amount of distillable entanglement is increased by the corner.
\item For holographic theories we have
\begin{equation}
\chi^{(4),{\rm E}}_{\rm sig}>\chi^{(4),{\rm E}}_{\rm sm},
\end{equation}
which may indicate that in the strong coupling limit, the amount of distillable entanglement is increased by the corners. This is the same as the $d=3$ case.
\item For complex scalars in different regimes of coupling,
\begin{equation}
  \chi^{(4),{\rm cs}}_{\rm sm}>\chi^{(4),{\rm E}}_{\rm sm},~~\chi^{(4),{\rm cs}}_{\rm sig}>\chi^{(4),{\rm E}}_{\rm sig},
\end{equation}
which means that in $d=4$, the monotonicity of $\chi$ (decreases as the coupling increases) is still preserved for singular entangling surfaces.
\item For free fermions,
\begin{equation}
\chi^{(4),{\rm f}}_{\rm sm}>\chi^{(4),{\rm E}}_{\rm sm},~~\chi^{(4),{\rm f}}_{\rm sig}>\chi^{(4),{\rm E}}_{\rm sig},
\end{equation}
which also respects the monotonicity, contrary to the $d=3$ case.
\end{itemize}

\subsection{Holographic considerations: Higher derivative gravity}
To see whether the higher derivative terms in gravity preserves the monotonicity of $\chi$, here we consider the general action~(\ref{qtg}).
As pointed out in~\cite{Myers:2010tj}, parameters in~(\ref{qtg}) should obey the following constraints arising from unitarity
\begin{eqnarray}
\lambda_{1}&=&-4\lambda_{2}=\lambda_{3}\equiv\frac{\lambda}{2},\nonumber\\
\mu_{4}&=&\frac{1}{32}(3\mu_{1}-12\mu_{2}-7\mu_{3}),\nonumber\\
\mu_{5}&=&\frac{1}{3}(-12\mu_{2}-5\mu_{3}),\nonumber\\
\mu_{6}&=&\frac{2}{27}(9\mu_{1}-48\mu_{2}-20\mu_{3}),\\
\mu_{7}&=&\frac{1}{72}(-45\mu_{1}+276\mu_{2}+97\mu_{3}),\nonumber\\
\mu_{8}&=&\frac{1}{216}(18\mu_{1}-96\mu_{2}-31\mu_{3}).\nonumber
\end{eqnarray}
In other words, the quadratic terms should take the form of Gauss-Bonnet gravity and the free parameters in the cubic terms are only $\mu_{1},\mu_{2},\mu_{3}$.

Consider the following ansatz for the black hole metric
\begin{equation}
ds^{2}=-\left(\frac{r^{2}}{L^{2}}f(r)-1\right)N(r)^{2}dt^{2}+\frac{dr^{2}}{\frac{r^{2}}{L^{2}}f(r)-1}+r^{2}d\Sigma_{3}^{2},
\end{equation}
it can be found that when
\begin{equation}
\mu_{1}=-\frac{1}{9}(60\mu_{2}+7\mu_{3}),
\end{equation}
the theory admits analytic solutions
\begin{eqnarray}
& &1-f(r)+\lambda f(r)^{2}+\tilde{\mu}f(r)^3=\frac{\omega^{4}}{r^{4}},~~N(r)=\frac{\tilde{L}}{R},\nonumber\\
& &\tilde{\mu}=-\frac{4}{9}(6\mu_{2}+\mu_{3}),~~\tilde{L}=\frac{L}{\sqrt{f_{\infty}}}.
\end{eqnarray}
Upon setting $\mu_{2}=0$, the above solution is reduced to the one in~\cite{Myers:2010tj} with $\mu_{3}=-9/4\tilde{\mu}$.
The black hole entropy is determined by Wald formula
\begin{equation}
S_{\rm BH}=\frac{A}{4G}(1+\lambda L^{2}S_{\rm GB}+\sum\limits_{i=1}^{8}\mu_{i}L^{4}S_{Z_{i}}),
\end{equation}
where $A=r_{H}^{3}V_{H}$ and
\begin{eqnarray}
S_{\rm GB}&=&2{R^{tr}}_{tr}-2(R^{t}_{t}+R^{r}_{r})+R,\nonumber\\
S_{Z_{1}}&=&3({R^{tm}}_{tn}{{R^{r}}_{rm}}^{n}-{R^{tm}}_{rn}{{R^{r}}_{tm}}^{n}),\nonumber\\
S_{Z_{2}}&=&6R^{trmn}R_{trmn},\nonumber\\
S_{Z_{3}}&=&2({R^{tr}}_{tm}{R_{r}}^{m}-{R^{tr}}_{rm}{R_{t}}^{m})+\frac{1}{2}(R_{mnpr}R^{mnpr}+R_{mnpt}R^{mnpt}),\nonumber\\
S_{Z_{4}}&=&4{R^{tr}}_{tr}R+R_{mnpq}R^{mnpq},\\
S_{Z_{5}}&=&R^{t}_{t}R^{r}_{r}-R^{t}_{r}R^{r}_{t}+{R^{r}}_{mrn}R^{mn}+{R^{t}}_{mtn}R^{mn},\nonumber\\
S_{Z_{6}}&=&\frac{3}{2}(R^{rm}R_{rm}+R^{tm}R_{tm}),\nonumber\\
S_{Z_{7}}&=&R_{mn}R^{mn}+R(R^{r}_{r}+R^{t}_{t}),\nonumber\\
S_{Z_{8}}&=&3R^{2}.\nonumber
\end{eqnarray}
The black hole entropy takes a simpler form after plugging in the metric
\begin{equation}
S_{BH}=\frac{A}{4G}(1-6\lambda f(r_{H})+9\tilde{\mu}f(r_{H})).
\end{equation}
Introducing $x\equiv r_{H}/L$, the temperature and entropy of the black hole read
\begin{equation}
T=\frac{1}{2\pi R}\frac{1}{x}\left(1+\frac{2}{f_{\infty}}\frac{x^{6}-f_{\infty}x^{4}+\lambda f_{\infty}^{2}x^{2}+\tilde{\mu}f_{\infty}^{3}}{x^{4}-2\lambda f_{\infty}x^{2}-3\tilde{\mu}f_{\infty}^{2}}\right),
\end{equation}
\begin{equation}
S_{\rm BH}=\frac{\tilde{L}^{3}V_{H}}{4G}x^{3}\left(1-6\lambda\frac{f_{\infty}}{x^{2}}+9\tilde{\mu}\frac{f_{\infty}^{2}}{x^{4}}\right).
\end{equation}

Hence the REs are given by~\cite{Hung:2011nu}
\begin{eqnarray}
S^{(4),{\rm H}}_{n}&=&\frac{\tilde{L}^{3}V_{H}}{8G}\frac{n}{n-1}(1-x_{n}^{2})\Big\{\frac{1+x_{n}^{2}}{f_{\infty}}+(1-16\lambda)-\frac{3\tilde{\mu}f_{\infty}^{2}}{x_{n}^{2}}-16f_{\infty}^{2}\times\nonumber\\
& &\left[\frac{((1-4\lambda)\lambda^{2}+3\tilde{\mu}\lambda-2\tilde{\mu})(x_{n}^{2}+3\tilde{\mu}f_{\infty}^{2})+3\tilde{\mu}f_{\infty}((1-2\lambda)\lambda+3\tilde{\mu})(1+x_{n}^{2}-2\lambda f_{\infty})}{(1-2\lambda f_{\infty}-3\tilde{\mu}f_{\infty}^{2})(x_{n}^{4}-2\lambda f_{\infty}x_{n}^{2}-3\tilde{\mu}f_{\infty}^{2})}\right]\Big\},\nonumber\\
\end{eqnarray}
where $x_{n}$ is determined by
\begin{equation}
\frac{2}{f_{\infty}}x_{n}^{6}-\frac{1}{n}x_{n}^{5}-x_{n}^{4}+\frac{2\lambda f_{\infty}}{n}x_{n}^{3}+\frac{3\tilde{\mu}f_{\infty}^{2}}{n}x_{n}-\tilde{\mu} f_{\infty}^{2}=0,
\end{equation}
with the constraint
\begin{equation}
1-f_{\infty}+\lambda f_{\infty}^{2}+\tilde{\mu}f_{\infty}^{3}=0.
\end{equation}
To obtain $\chi$ we need the EE
\begin{equation}
\lim_{n\rightarrow1}S^{(4),{\rm H}}_{n}=\frac{\tilde{L}^{3}V_{H}}{4G}(1-6\lambda f_{\infty}+9\tilde{\mu} f_{\infty}^{2}),
\end{equation}
and the scaling dimension of twist operator~\cite{Hung:2011nu}
\begin{equation}
h^{(4),{\rm H}}_{n}=\frac{\tilde{L}^{3}}{8G}\frac{n(1-x_{n}^{2})(x_{n}^{2}+x_{n}^{4}-x_{n}^{2}f_{\infty}-\tilde{\mu}f_{\infty}^{2})}{x_{n}^{2}f_{\infty}},
\end{equation}
as well as $\partial_{n}h_{n}|_{n=1}$
\begin{equation}
\partial_{n}h_{n}|_{n=1}=\frac{\tilde{L}^{3}}{12G}(1-2\lambda f_{\infty}-3\tilde{\mu}f_{\infty}^{2}).
\end{equation}

The general expression for $\chi$ is very complicated, so we have to work out the perturbative solution at leading order in $\lambda$ and $\tilde{\mu}$. It can be derived that at $n=1/2$, $x_{n}$ has the following expansion
\begin{equation}
x_{1/2}=\frac{1}{2}(1+\sqrt{3})+\frac{1}{2}(2\sqrt{3}-3)\lambda+\frac{1}{2}(30\sqrt{3}-51)\tilde{\mu},
\end{equation}
which leads to the results for $\chi$
\begin{equation} \label{approx1}
\chi^{(4),{\rm H}}_{\rm sm}=\frac{S^{(4),{\rm H}}_{1/2}}{S^{(4),{\rm H}}_{1}}\approx1.67404+3.375\lambda-6.4702\tilde{\mu},
\end{equation}
\begin{equation} \label{approx2}
\chi^{(4),{\rm H}}_{\rm sig}=\frac{\sigma^{(4),{\rm H}}_{1/2}}{\sigma^{(4),{\rm H}}_{1}}\approx2.42404+3.72308\lambda+8.23557\tilde{\mu}.
\end{equation}
At first sight, the monotonicity may be broken or preserved, depending on the values of $\lambda$ and $\tilde{\mu}$. To further explore this issue we resort to numerics and for simplicity we
take $\mu_{2}=0,~~\tilde{\mu}=\mu$, so the black hole metric becomes the one studied in~\cite{Myers:2010jv}.  In addition, we have to impose the following constraints arising from positivity of energy fluxes~\cite{Myers:2010jv}
\begin{eqnarray}
\label{lmmu}
0\leq1-10\lambda f_{\infty}+189\mu f_{\infty}^2,\nonumber\\
0\leq1+2\lambda f_{\infty}-855\mu f_{\infty}^{2},\\
0\leq1+6\lambda f_{\infty}+1317\mu f_{\infty}^{2}.\nonumber
\end{eqnarray}
We plot the difference between $\chi^{(4),{\rm H}}_{\rm sm}$ and $\chi^{(4),{\rm H}}_{\rm sig}$ in the permitted parameter range~(\ref{lmmu}), aiming at checking whether singular entangling surface would increase the amount of distillable entanglement. 
\begin{figure}[t]
    \centering
    \includegraphics[width=0.8\textwidth]{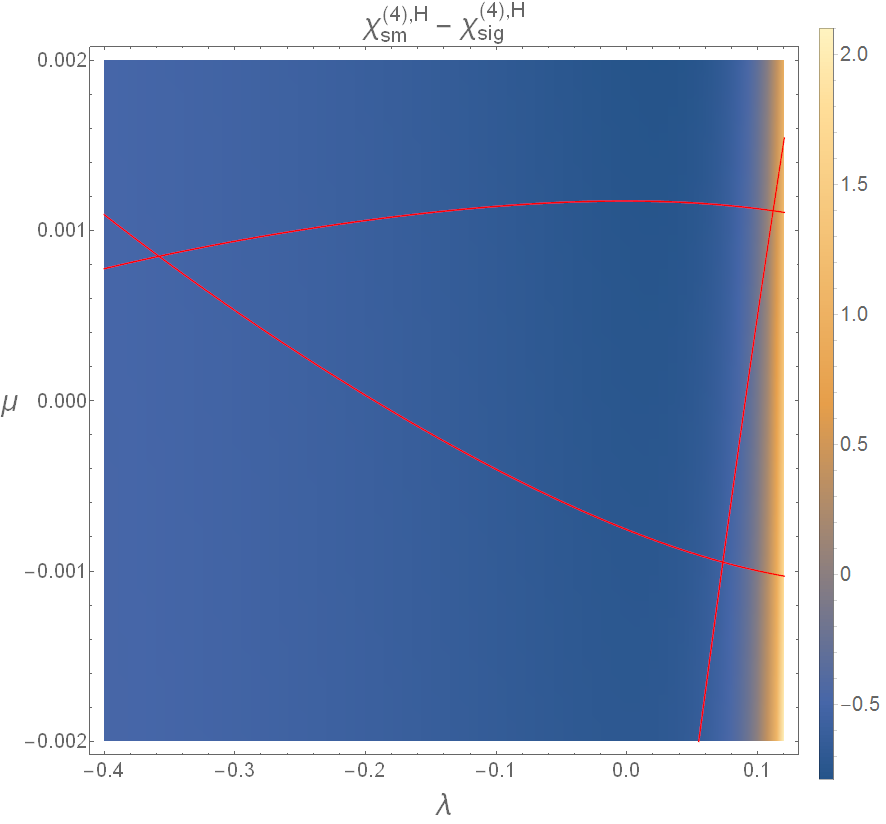}
    \caption{The difference between $\chi^{(4),{\rm H}}_{\rm sm}$ and $\chi^{(4),{\rm H}}_{\rm sig}$ in the permitted parameter range~(\ref{lmmu}) surrounded by the three red curves. {The region enclosed by three red curves satisfies the three inequalities \eqref{lmmu}.}}
    \label{fig1}
\end{figure}
From~Fig.~\ref{fig1} we may conclude that in most of the permitted parameter range (the area surrounded by the three red curves), the difference between $\chi^{(4),{\rm H}}_{\rm sm}$ and $\chi^{(4),{\rm H}}_{\rm sig}$ is negative, which would suggest that the amount of distillable entanglement increases for singular entangling regions within most of the parameter range. However, there exists a small region near the boundary of the allowed parameter range, where $\chi^{(4),{\rm H}}_{\rm sm}$ is larger than $\chi^{(4),{\rm H}}_{\rm sig}$. This may be seen as an indication of decrease of distillable entanglement for singular entangling surfaces in that regime.
A more explicit exhibition of the `distillable entanglement decreasing' region is given as follows.
\begin{figure}[t]
    \centering
    \includegraphics[width=0.8\textwidth]{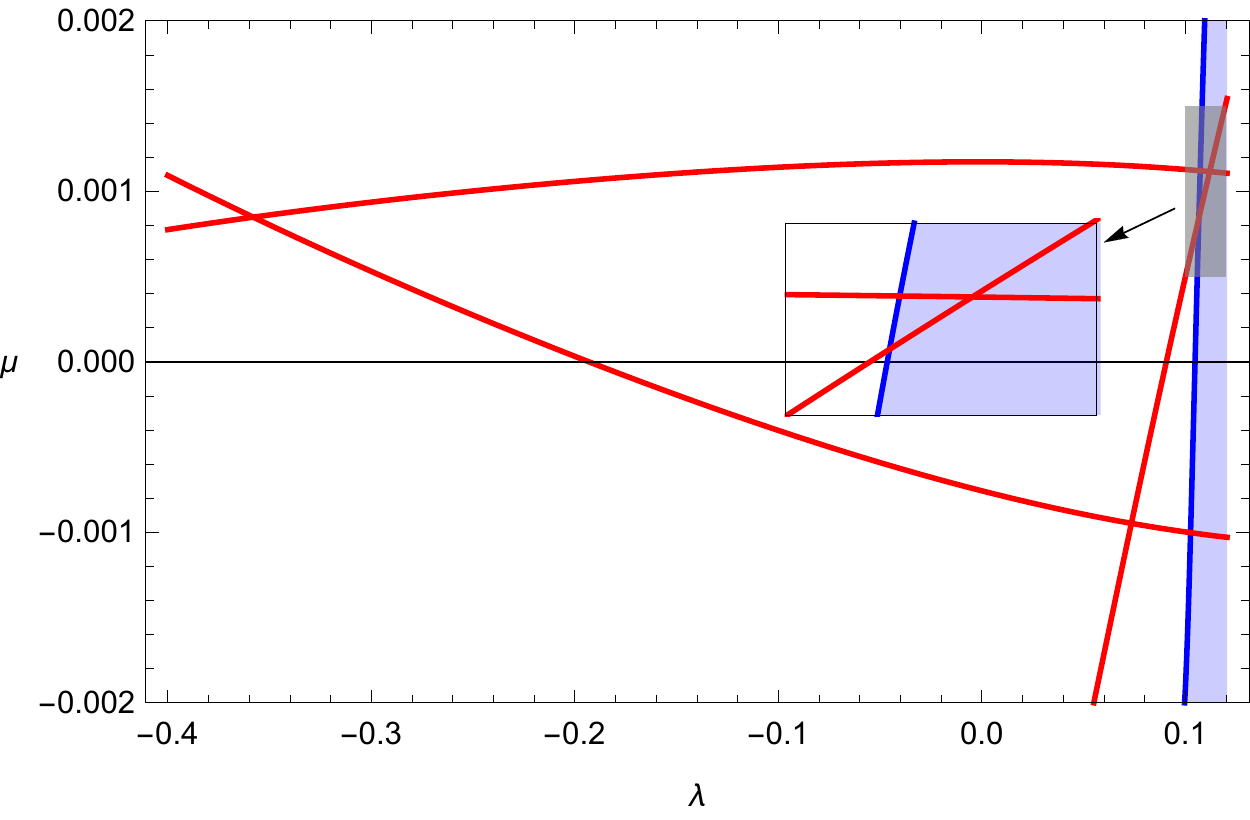}
    \caption{The region where the amount of distillable entanglement decreases for singular entangling regions: the small triangle region surrounded by one blue line and two red curves in the subplot. {The blue line designates the parameters for  $\chi^{(4),{\rm H}}_{\rm sm}-\chi^{(4),{\rm H}}_{\rm sig}=0$ and the region enclosed by three red curves satisfies the three inequalities \eqref{lmmu}. }}
    \label{fig2}
\end{figure}
 \begin{figure}[t]
    \centering
    \includegraphics[width=1\textwidth]{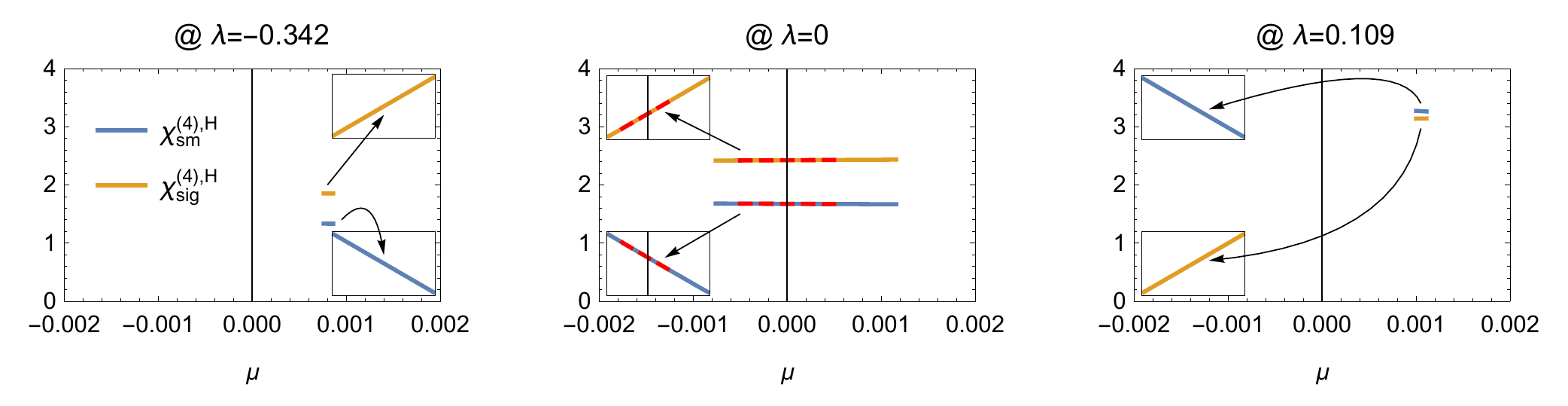}
    \caption{$\chi^{(4),{\rm H}}_{\rm sm}$ and $\chi^{(4),{\rm H}}_{\rm sig}$ at fixed $\lambda$. {All lines in the figures look constant because of the scale difference. To see they increase or decrease, we put the insets, where the vertical axis was scaled up.  The dotted lines in the inset of the middle figure express \eqref{approx1} and \eqref{approx2} for small $\mu$ at $\lambda=0$. }} 
    \label{fig3}
\end{figure}
\begin{figure}[t]
    \centering
    \includegraphics[width=1\textwidth]{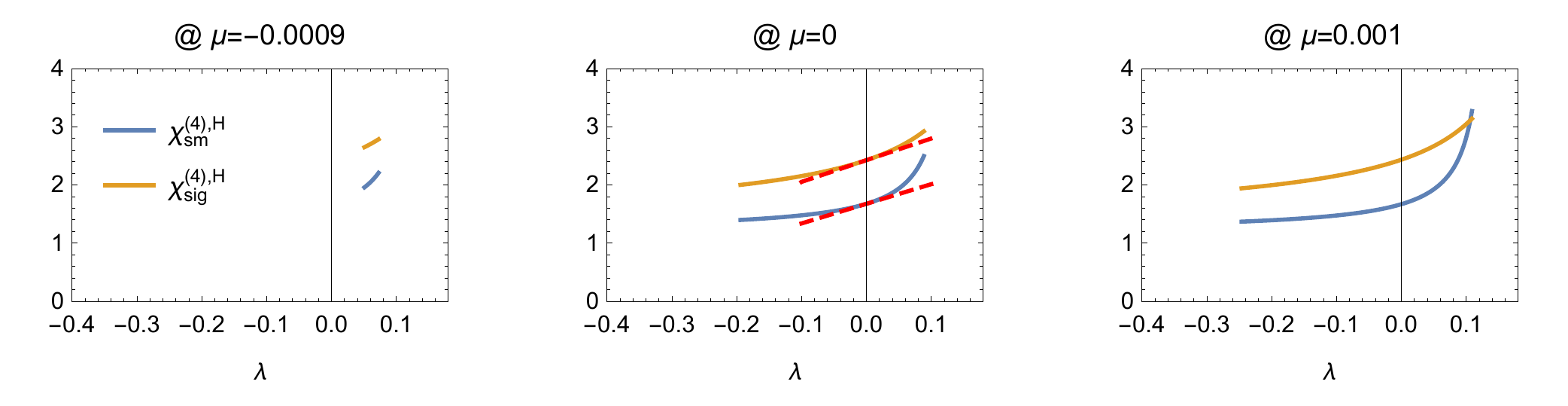}
    \caption{$\chi^{(4),{\rm H}}_{\rm sm}$ and $\chi^{(4),{\rm H}}_{\rm sig}$ at fixed $\mu$. {The dotted lines in the middle figure express \eqref{approx1} and \eqref{approx2} for small $\lambda$ at $\mu=0$.  } 
    }
    \label{fig4}
\end{figure}

To explicitly explore how the amount of distillable entanglement evolves as the parameters vary, in Fig. \ref{fig3} and \ref{fig4}, we plot $\chi^{(4),{\rm H}}_{\rm sm}$ and $\chi^{(4),{\rm H}}_{\rm sig}$ at some fixed representative values of $\mu$ or $\lambda$, while the other parameter is varying. For all the cases $\lambda$ and $\mu$ should take values in the range determined by~(\ref{lmmu}). From these plots we can observe

\begin{itemize}
\item $\chi^{(4),{\rm H}}_{\rm sig}>\chi^{(4),{\rm H}}_{\rm sm}$ for most of the cases, while $\chi^{(4),{\rm H}}_{\rm sig}<\chi^{(4),{\rm H}}_{\rm sm}$ when the parameters take values in the `distillable entanglement decreasing' region, as shown in the right plot of both figures.
\item {At fixed $\lambda$ (Fig. \ref{fig3}), both $\chi^{(4),{\rm H}}_{\rm sig}$  monotonically increase but $\chi^{(4),{\rm H}}_{\rm sm}$ monotonically decrease as $\mu$ increases, while at fixed $\mu$ (Fig. \ref{fig4}), both $\chi^{(4),{\rm H}}_{\rm sig}$ and $\chi^{(4),{\rm H}}_{\rm sm}$ monotonically increase as $\lambda$ increases. The dotted lines in the middle plots of both figures represent the approximate formula \eqref{approx1} and \eqref{approx2} for small $\mu$ and $\lambda$. } 
\item The infinite coupling limit corresponds to taking $\lambda=\mu=0$. As seen from the plots, the amount of distillable entanglement may not necessarily increase as one turns on finite coupling, that is, nonzero $\lambda$ or $\mu$. The tendency depends on the sign of the parameters.
\end{itemize}

\section{Summary and discussion}

Recent advances in gauge/gravity enable us to evaluate entanglement measures such as EE, RE and EN in terms of geometric quantities, thanks to the Ryu-Takayanagi formula and various generalisations. Among all these quantities the EN has been much less understood, perhaps because of the notorious square root in the definition~(\ref{osq}). However, for pure states of a bipartite system the evaluation can be largely simplified. As shown in~\cite{Calabrese:2012nk}, the EN is equal to the RE with index $1/2$, which allows us to compute the EN both from field theory perspective and holographically. Furthermore, the authors of~\cite{Rangamani:2014ywa} defined a universal, cutoff independent quantity $\chi$ given by the ratio of the universal part of the EN and the EE and argued that $\chi$ characterises the amount of distillable entanglement. They calculated $\chi$ in $d=3$ free CFTs and $d=4$ SYM, as well as their gravity duals and observed that $\chi$ takes a larger value in free theories, which may suggest that the amount of distillable entanglement decreases as the coupling constant increases. They further claimed that $\chi$ should depend only on the geometry of the entangling surface and later in~\cite{Perlmutter:2015vma} they found that the monotonicity, that is, $\chi$ monotonically decreases as the coupling constant grows, could be violated if the entangling region contains higher genus.

\begin{center}
	\begin{table}[t]
		\hspace{0.1cm}
		\begin{tabular}{|c||c|c||c|c|}
			\hline
			&
			\multicolumn{2}{c||}{Free field theory}&
			\multicolumn{2}{c|}{Holography} 
			\\ \hline
			&
			Complex scalar & Fermion&
			$N\rightarrow \infty$ & Higer derivative correction 
			\\ \hline 
			$d=3$& 
			$\chi^{(3),{\rm cs}}_{\rm sm}=2.71579$& 
			$\chi^{(3),{\rm f}}_{\rm sm}=1.88752$&
			$\chi^{(3),{\rm H}}_{\rm sm}\approx1.63113$&$-3.05088\lambda_{3}$
			\\ \hline
			$d=3$& 
			$\chi^{(3),{\rm cs}}_{\rm sig}\approx2.54648$& 
			$\chi^{(3),{\rm f}}_{\rm sig}\approx2.16152$&
			$\chi^{(3),{\rm H}}_{\rm sig}\approx2.16509$&$-2.40478\lambda_{3}$
			\\ \hline
			$d=4$& 
			$\chi^{(4),{\rm cs}}_{\rm sm}=3.75$& 
			$\chi^{(4),{\rm f}}_{\rm sm}\approx2.21591$&
			$\chi^{(4),{\rm H}}_{\rm sm}\approx1.67404$&$+3.375\lambda-6.4702\tilde{\mu}$
			\\ \hline
			$d=4$& 
			$\chi^{(4),{\rm cs}}_{\rm sig}=3.75$& 
			$\chi^{(4),{\rm f}}_{\rm sig}=2.8125$&
			$\chi^{(4),{\rm H}}_{\rm sig}\approx2.42404$&$+3.72308\lambda+8.23557\tilde{\mu}$
			\\ \hline
		\end{tabular}
		\caption{Summary of $\chi$'s. The rightmost column summrizes the approximate formulas for small parameters ($\lambda_3, \lambda, \tilde{\mu}$). For full parameter region for $d=4$, see Fig. \ref{fig3} and \ref{fig4}. }
		\label{table:sum}
	\end{table}
\end{center}

\vspace{-1.7cm}

In this paper we consider the contribution to $\chi$ from a relatively simple class of entangling surfaces: surfaces containing corners. 
Our results are summarized in Table \ref{table:sum}. 
The presence of corners would induce new universal contributions to the EE and RE, as studied in~\cite{Bueno:2015rda, Bueno:2015xda}. Such new universal contributions are proportional to $\log H/\epsilon$ for odd $d$ and to $\log^{2}H/\epsilon$ for even $d$ and the proportional coefficients are generically determined by the scaling dimensions of twist operators. We calculate $\chi$ in $d=3$ and $d=4$ for free complex scalars and free Dirac fermions, and compare the results with their smooth counterparts. For complex scalars in $d=3$, the amount of distillable entanglement decreases in the presence of corners, while in $d=4$ it remains invariant (the first column of Table \ref{table:sum}). However, the amount of distillable entanglement increases for free fermions with singular entangling regions in both $d=3$ and $d=4$ (the second column of Table \ref{table:sum}). In the strong coupling limit we compute $\chi$ holographically in Einstein gravity and find that $\chi$ increases for singular entangling surfaces in both $d=3$ and $d=4$ (the third column of Table \ref{table:sum}).

To explore the monotonicity of $\chi$ for singular entangling surfaces, we first compare $\chi$ evaluated from free theories and Einstein gravity. We find that for complex scalars the monotonicity is still preserved for singular surfaces in both $d=3$ and $d=4$, while for free Dirac fermions the monotonicity is preserved in $d=4$ and is violated (although by a small amount) in $d=3$ in the presence of corners. Our observation may be seen as another evidence that the monotonicity could be affected by the geometry of the entangling surface. However, another interpretation of this violation might be that the gravity dual of free Dirac fermions cannot be simple Einstein theory but rather Vasiliev's higher spin theory, so to consistently evaluate $\chi$ one should consider Vasiliev gravity in AdS instead and the monotonicity might be respected in the new framework.

We also study finite $N$/finite-coupling effects to $\chi$ in the holographic setup by considering higher derivative gravity theories (the forth column of Table \ref{table:sum}, Fig. \ref{fig3} and \ref{fig4}). In $d=3$ the higher derivative action contains the squares of Ricci tensor and Ricci scalar, as well as the cubic term of Weyl tensor. Such a theory does not admit exact topological black hole solutions for computing the scaling dimension of twist operators and we work in perturbations of the parameters $\lambda_{i}$ in front of the higher derivative terms. Although the RE with index $1/2$ contains all the three $\lambda_{i}$s, the resulting $\chi$ only depends on the coefficient of the Weyl tensor. Therefore the amount of distillable amount could either increase or decrease, depending on the sign of the parameter. At present we cannot fix the allowed range of $\lambda_{i}$ as we are working in perturbative expansions. In $d=4$ we consider quasi-topological gravity~\cite{Myers:2010ru, Myers:2010jv}, which allows for exact topological black hole solutions. From our numerical plots we can see that in most of the allowed parameter region, the amount of distillable entanglement increases for singular entangling surfaces, but there does exist a small regime where it decreases. Moreover, $\chi$ grows monotonically as one fixes one parameter in quasi-topological gravity and varies the other. The results may also indicate that as we incorporate finite $N$/finite-coupling effects, the amount of distillable entanglement depends on the sign of the relevant parameter.

Entanglement negativity gives an upper bound on distillable entanglement so it is expected to be greater than entanglement entropy. 
It suggests that $\chi > 1$. However,
it was shown~\cite{Perlmutter:2015vma} that  $\chi$ can be less than one if the geometry and topology of the entangling surface are
complicated enough. From our computation we see that $\chi > 1$ always also for the singular entangling surface.

One might conclude that the presence of corner contributions would increase the amount of distillable entanglement from most of the examples we have studied. However, we have also found several counterexamples: $\chi$ for free complex scalars in $d=3,4$ and for quasi-topological gravity in $d=4$ in a small parameter region. In these cases $\chi$ takes a smaller value for singular entangling regions. These observations would raise the following questions: How will the geometry of the entangling region affect distillable entanglement? To what extent singular entangling surfaces would increase distillable entanglement? It would be very interesting to study these questions at a more general level. In particular, how would our conclusions change for mixed states? However to our knowledge concrete analysis for EN in mixed states was only performed for $d=2$ CFTs in~\cite{Calabrese:2012nk}, while holographic computations have not been available. We hope to address these questions in the future.  

\bigskip \goodbreak \centerline{\bf Acknowledgments}
\noindent

DWP would like to thank Andy O'Bannon for collaborations at an early stage of this project, very helpful discussions on related topics and valuable comments on the manuscript. He would also like to thank Pablo Bueno, Rom\'{a}n Or\'{u}s and Erik Tonni for discussions.
DWP is supported by the Marie Curie Intra-European Fellowship nr 622730 within the 7th European Community
Framework Programme FP7/2007-2013.
The work of KYK and CN was supported by Basic Science Research Program through the National Research Foundation of Korea(NRF) funded by the Ministry of Science, ICT \& Future Planning(NRF-2014R1A1A1003220) and the GIST Research Institute(GRI) in 2016.



\end{document}